\documentclass[conference]{IEEEtran}
\IEEEoverridecommandlockouts
\usepackage{cite}
\usepackage{comment}
\usepackage{amsmath,amssymb,amsfonts}
\usepackage{algorithmic}
\usepackage{graphicx}
\usepackage{textcomp}
\usepackage{xcolor}
\usepackage{url}
\def\BibTeX{{\rm B\kern-.05em{\sc i\kern-.025em b}\kern-.08em
    T\kern-.1667em\lower.7ex\hbox{E}\kern-.125emX}}
\begin{document}

\title{Enhanced Accessibility for\\ Mobile Indoor Navigation\\
\thanks{The work presented in this paper was supported by the German Federal Ministry for Digital and Transport (BMDV).}
}

\author{\IEEEauthorblockN{Johannes Wortmann, Bernd Schäufele}
\IEEEauthorblockA{\textit{Fraunhofer Institute for Open}\\
\textit{Communication Systems (FOKUS)} \\
Berlin, Germany \\
\{johannes.wortmann,\\bernd.schaeufele\}@fokus.fraunhofer.de
}
\and
\IEEEauthorblockN{Konstantin Klipp, Ilja Radusch}
\IEEEauthorblockA{\textit{Daimler Center for Automotive} \\
\textit{IT Innovations (DCAITI)}\\
Berlin, Germany \\
\{konstantin.klipp,\\ilja.radusch\}@dcaiti.com}
\and
\IEEEauthorblockN{Katharina Blaß, Thomas Jung}
\IEEEauthorblockA{\textit{Hochschule für Technik und Wirtschaft} \\
\textit{University of Applied Sciences}\\
Berlin, Germany \\
me@katharinablass.dev\\
thomas.jung@htw-berlin.de}
}

\maketitle

\begin{abstract}

The navigation of indoor spaces poses difficult challenges for individuals with visual impairments, as it requires processing of sensory information, dealing with uncertainties, and relying on assistance. To tackle these challenges, we present an indoor navigation app that places importance on accessibility for visually impaired users. Our approach involves a combination of user interviews and an analysis of the Web Content Accessibility Guidelines. With this approach, we are able to gather invaluable insights and identify design requirements for the development of an indoor navigation app. Based on these insights, we develop an indoor navigation app that prioritizes accessibility, integrating enhanced features to meet the needs of visually impaired users. The usability of the app is being thoroughly evaluated through tests involving both visually impaired and sighted users. Initial feedback has been positive, with users appreciating the inclusive user interface and the usability with a wide range of accessibility tools and Android device settings. 

\end{abstract}


\section{Introduction}
Addressing challenges faced by visually impaired individuals is an important, yet challenging, task for the society. Visual impairments not only impact the individual, but also families, communities, and societies as a whole. Barriers to independence and mobility in indoor spaces can limit daily lives. Challenges include difficulties in unfamiliar environments, reliance on others, and limited access to services.

Accessible indoor navigation is an important element to address these challenges. Such an app can greatly improve the quality of life for visually impaired individuals, giving them independence and quality of life in indoor spaces. It enables access to education, employment, healthcare, and social activities, promoting inclusion and equal opportunities.

The field of indoor navigation has advanced significantly and reached product level, e.g., with the everGuide app \cite{everguideWeb} by Fraunhofer FOKUS providing precise navigation in complex indoor settings. This paper covers the concept and development of a new version of this app that has enhanced accessibility features, specifically optimized for visually impaired users. We aim to improve the accessibility features with a systematic methodology, incorporating user feedback from interviews and testing, and directly surveying the target group of blind and visually impaired users. Our work focuses on a user-centered design addressing navigation accuracy, user interface (UI) design, compatibility with assistive technologies, feedback mechanisms, and smartphone operation requirements. 

These efforts target visually impaired users' unique needs, ultimately enhancing their mobility, independence, and inclusivity in navigating indoor spaces. This entails not only user requirements for indoor navigation but also the missing requirements for smartphone operation and support for assistive systems and settings, as identified in related works. Directly surveying the target group of blind and visually impaired users is suitable for this purpose. However, recruiting a sufficient number of participants representing various types of visual impairments can be challenging \cite{xie2022assessment}, \cite{guarese2022evaluating}. Thus, this paper also incorporates the WCAG accessibility standards \cite{Henry2023} during the development of the optimized version of the everGuide App. Furthermore, the UI concept applies the Material Design \cite{mew2015learning}.

The paper is structured as follows. Section \ref{sec:related_work} gives an overview of relevant research in this field. 
Subsequently, the realization in the navigation app follows in Section \ref{sec:app}. The developed app is evaluated, and the results are discussed and interpreted in Section \ref{sec:evaluation}. Finally, a conclusion and outlook on future work are given in section \ref{sec:summary}.

\section{Related work}
\label{sec:related_work}
Extensive research is conducted in the field of navigation systems for visually impaired users, addressing various aspects such as turn-by-turn navigation \cite{ahmetovic2016navcog, sato2019navcog3, prudtipongpun2015indoor, khusro2022haptic}, mental map generation of the surroundings \cite{paratore2023exploiting}, \cite{aziz2022planning}, and obstacle detection \cite{presti2019watchout} with and without additional devices such as belts \cite{cosgun2014evaluation, pawar2022smartphone}. Particularly relevant are user requirements, UI design, and the communication modality of feedback during navigation. 

In \cite{jeamwatthanachai2019indoor} a study is conducted on the navigation behavior and challenges faced by blind and visually impaired users in unfamiliar buildings. The insights from semi-structured interviews and questionnaires with the target group and experts emphasize the users' uncertainty in such environments. These findings significantly contribute to the understanding of the target audience and represent a comprehensive foundation for further research. Though, there are several relevant topics missing in this study, particularly regarding smartphone usage and the associated requirements for a navigation application.

In another study 614 blind users are surveyed in an empirical investigation regarding the necessary feedback information and features of a navigation app \cite{ponchillia2020developing}. The results emphasize the communication of Points of Interest (POI), e.g., elevators and toilets, as well as room numbers during indoor navigation. Users also consider the real-time location information and the ability to create routes to selected destinations as particularly relevant. These are identified as the most important features, followed by announcements of POIs on the way and the customization of navigation feedback in terms of type and level of detail. This study is highly relevant for the development of a navigation application focusing on blind users. However, neither the requirements for smartphone operation nor the available assistance systems are considered in both studies.

Most research in navigation systems for blind and visually impaired users utilizes tailored graphical user interfaces (GUIs) with simple designs and large control elements \cite{ahmetovic2016navcog}, \cite{prudtipongpun2015indoor, khan2022mechanism}. Another approach, discussed by Khan and Khusro \cite{khan2022mechanism}, is Adaptive Design, enabling customization of interfaces for blind users and compatibility with external devices like smartwatches. This approach is relevant for developing different versions of the navigation app to accommodate various types of visual impairments in the target group.

\begin{figure}[bt!]
    \centering
        \includegraphics[width=1\columnwidth]{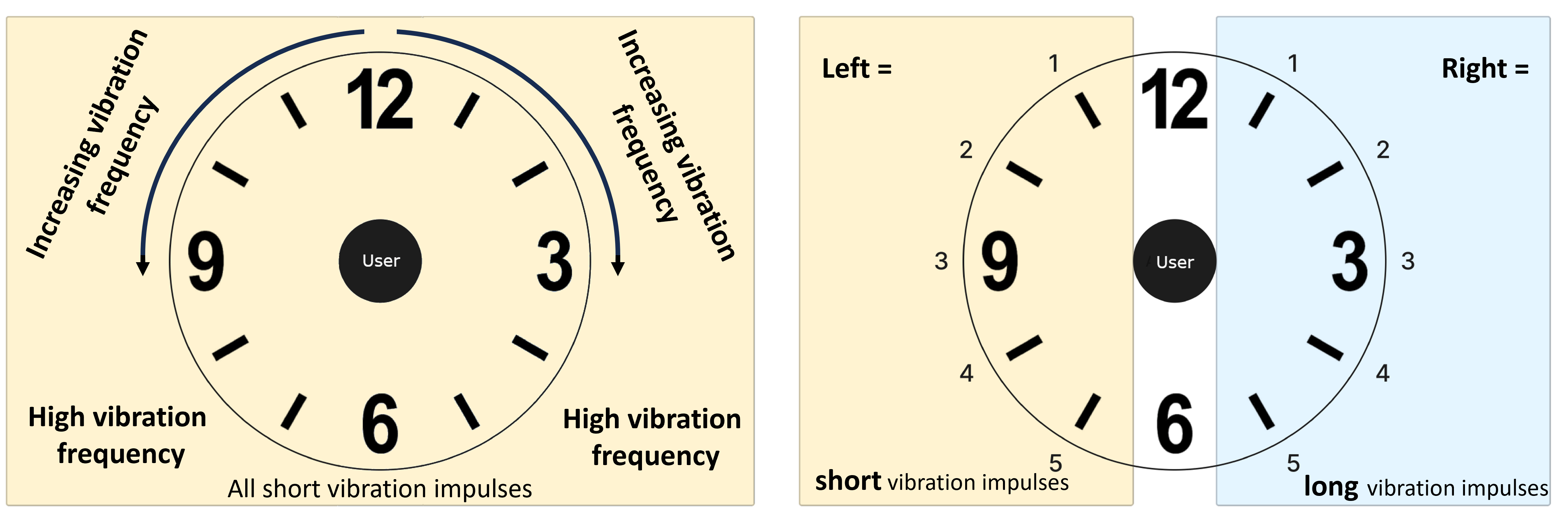}
    \caption{Vibration pattern for direction encoding: Compass-Style for continuous encoding (left) and Counting-Clock-Style for event-based encoding (right)}
    \label{fig:direction}
\end{figure}

A key aspect in the development of a navigation application is the design of navigation feedback to convey necessary information regarding distances, changes in direction, and obstacles along the route. In related studies, various approaches have been explored to implement this feedback, including voice guidance \cite{ahmetovic2016navcog, sato2019navcog3, prudtipongpun2015indoor, paratore2023exploiting}, non-verbal sounds (sonification) \cite{ahmetovic2016navcog, sato2019navcog3, presti2019watchout}, and vibration \cite{khusro2022haptic, paratore2023exploiting, cosgun2014evaluation, pawar2022smartphone}.

Despite a tendency towards speech as feedback medium in these studies, there is no clear superiority of a specific feedback type. Instead, all mentioned studies provide relevant design approaches for our work. Another relevant research \cite{ohn2018variability} focuses on the timing of navigation feedback to avoid error situations, e.g., overshooting during turns. Factors influencing blind users' response time to navigation feedback include walking speed and individual navigation style.

\section{Indoor Navigation App}
\label{sec:app}

The everGuide SDK \cite{everguideWeb} forms the core of our application, providing precise positioning, route generation and navigation functionality. 
To achieve the required level of precision, everGuide combines the approach from \cite{klipp2018rotation} with optical markers. This allows us to fully focus our development on the usability aspect of creating an app for visually impaired users.

\subsection{Navigation feedback concept}
\label{sec:concept}

The everGuide app originally uses voice announcements and an acoustic compass for navigation feedback. The acoustic compass uses clicking sounds for navigation feedback, with the pace of clicks corresponding to the degree of directional deviation. This implies that the absence of noises indicates precise alignment with the route, whereas the fastest rhythm of the click sounds signifies a maximum deviation of 180 degrees. This concept is visualized in Fig. \ref{fig:direction} ("Compass-Style").

Acoustic feedback, also known as sonification, utilizes non-verbal sounds to encode information, offering privacy benefits \cite{khusro2022haptic}. It provides various parameters such as pitch, timbre, volume, spatiality, and timing, and can even incorporate musical elements like tonality\cite{ahmetovic2019sonification, dubus2013systematic}. Some studies use auditory icons for sonification, representing real objects with sounds, like bell tones for a church \cite{aziz2022planning}. However, the wide range of adjustable parameters can make acoustic patterns complex and hard to interpret, and they may be poorly perceived in noisy environments, similar to voice announcements \cite{khusro2022haptic}.

Haptic feedback with vibrations is effective when visual and auditory senses are unavailable, language-independent, and ensures privacy \cite{kammoun2012guiding}. However, it can increase cognitive load when users need to distinguish multiple patterns. Moreover, vibration patterns have fewer adjustable parameters: frequency, pulse length, and rhythm, compared to sonification \cite{khusro2022haptic}.

Comparing vibration and acoustic feedback indicates that vibration could more effectively complement voice announcements in the app, as it lacks the disadvantages of audio. To test this, a feedback concept with both vibration and acoustic versions is needed, where users can compare both directly. The existing acoustic compass concept is adapted to include a vibration version. The design of the vibration compass is identical to the acoustic compass, using vibration pulses instead of clicking sounds and increasing the vibration rhythm as directional deviation increases.

\begin{figure}[hbt!]
    \centering
    \includegraphics[width=1\columnwidth]{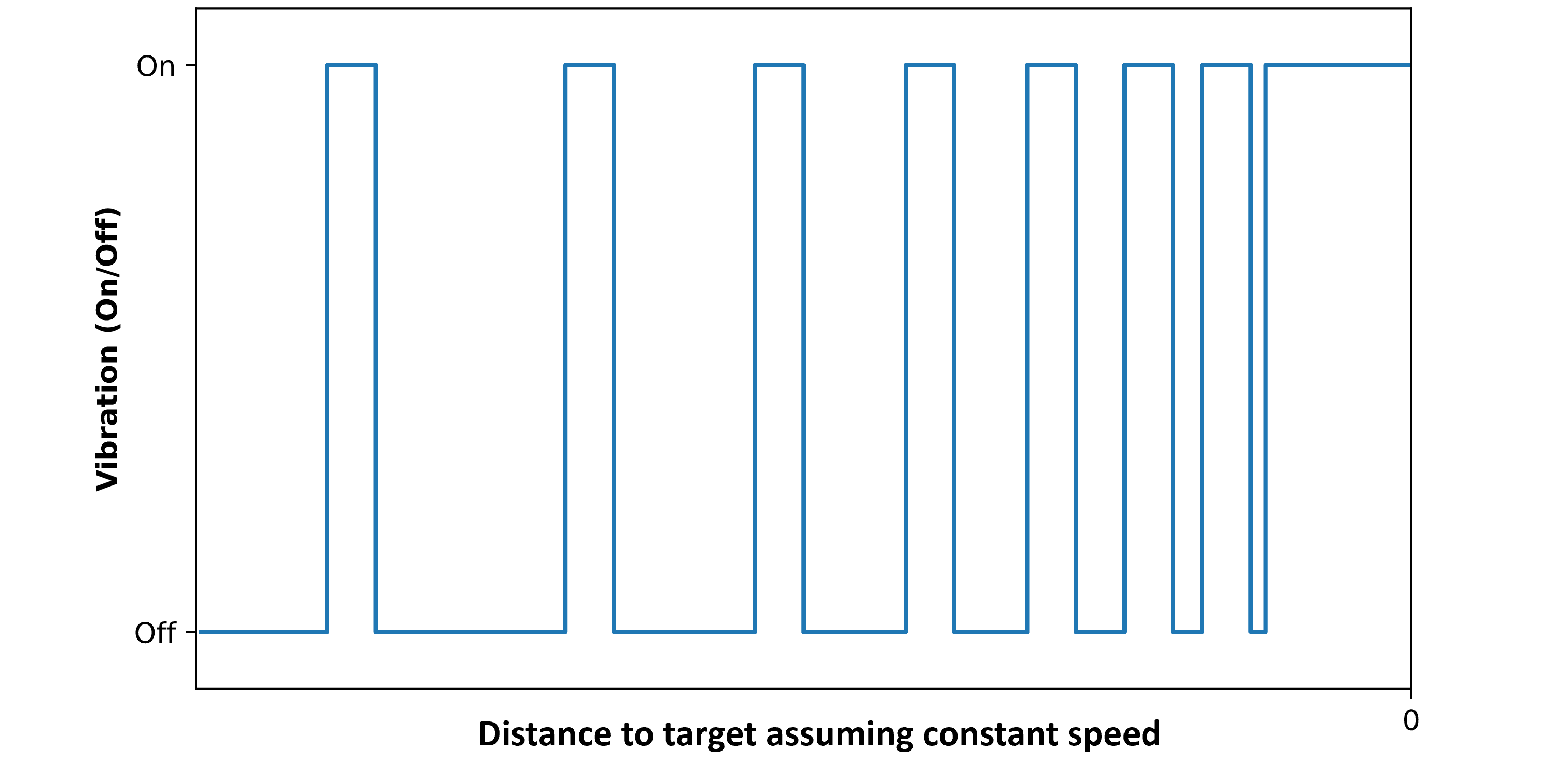}
        \caption{Vibration pattern for distance encoding: The closer we get to the target, the higher the vibration frequency.}
    \label{fig:distance}
\end{figure}

\begin{figure*}[hbt!]
    \centering
    \includegraphics[width=0.7\textwidth]{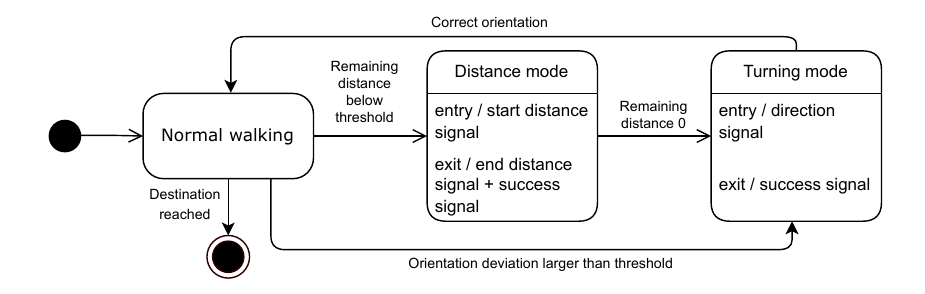}
    \caption{Navigation workflow for event-based navigation}
        \label{fig:state}
\end{figure*}

The compass might be considered too distracting due to its continuous feedback, rather than only at relevant events like reaching a turn. Therefore, an event-based feedback concept is developed and compared to the continuous compass feedback. Both concepts use the same feedback medium to avoid bias, with the vibration compass suggested for comparison.

 Different navigation information can be communicated through vibration, e.g., status updates, distance and direction at junctions, floor change, and obstacle warnings. However, not all information can be easily encoded for quick responses, e.g., obstacle warnings. Moreover, too many different information types and corresponding vibration patterns can complicate understanding \cite{khusro2022haptic, paratore2023exploiting}, causing increased cognitive load. Thus, only key details like distance and direction are conveyed by vibration, with other information still provided via voice for faster interpretation.

Vibration patterns can be created by modulating the frequency, duration, and interval of pulses, similar to car parking sensors where the beep speed inversely indicates the distance to obstacles. This concept, adapted for indoor navigation to communicate distances via vibrations \cite{ahmetovic2016navcog, presti2019watchout}, starts with slow pulses and increases frequency as the user nears a turn, using pulse intervals to indicate remaining distance (See Fig. \ref{fig:distance}). This analogy to parking sensors makes the vibration pattern easy to learn within the app. The new concept includes encoding directional information at junctions, originally communicated through voice prompts like "turn slightly right." Besides junctions, the directional information is used when the user deviates slightly from the correct orientation.

The first method for communicating directions, option A, separates the information about turning direction and angle. The direction is indicated by the length of the vibration pulses, and the angle by the number of pulses. This approach uses a clock system, indicating directions by clock positions such as 2 o'clock for "turn slightly right." For example, a sequence of three long pulses signifies a right turn to 3 o'clock, while three short pulses mean a left turn to 9 o'clock. This allows precise direction with up to 6 pulses but increases complexity as users must interpret the number and length of pulses. This option is depicted in Fig. \ref{fig:direction} as "Counting-Clock-Style".

Option B for direction guidance uses the ping method from Ahmetovic et al.\cite{ahmetovic2019sonification}, providing a single vibration pulse to prompt users to rotate until correctly oriented. This simpler approach does not specify the turn degree or direction, reducing complexity, but also precision in navigation. In the proposed direction encoding methods, users need a signal to indicate when their directional adjustment is complete. Such a signal could be a single, notably long vibration pulse that activates once the user is correctly oriented.

Two versions of event-based vibration are integrated into the app. The first one uses option B for direction encoding and the introduced concept for distance. The second one combines the same distance concept with the more complex option A (Counting-Clock-Style) for direction signals. Both versions require clear differentiation between direction and distance indications, achieved via the previously mentioned completion signal played after each distance reached and direction change. The states of the event-based guidance are shown in Fig. \ref{fig:state}.

\subsection{User interface design}
\label{sec:ui}
The app design process is structured into two phases using the design tool Figma \cite{staiano2022designing}, involving multiple iterations.
The first phase focuses on creating and comparing several low-fidelity designs, which emphasize the layout and user interaction flow without specific texts, graphics, or colors. 
This phase aims to develop an app structure that is simple, intuitive and adhering to the ”Design4All” principle for both sighted and visually impaired users. 
The most suitable low-fidelity designs are refined into high-fidelity versions in the second phase, adding all necessary details such as texts and colors for a complete app design.

For the high-fidelity prototype, user interface components can be custom-built or integrated from existing libraries like Material Design \cite{mew2015learning}. It provides design guidelines and UI components for interfaces based on best practices and accessibility standards. Using Material Design components, offers compatibility with new Android apps and ensures component-level accessibility. The latest version, Material Design 3, features enhancements like larger UI components complying with the WCAG standard and is default on Android devices from version 12. Due to its benefits and alignment with the "Design4All" concept, it is selected as the design for the new prototype.

\begin{figure}[b!]
    \centering
    \includegraphics[width=0.7\columnwidth]{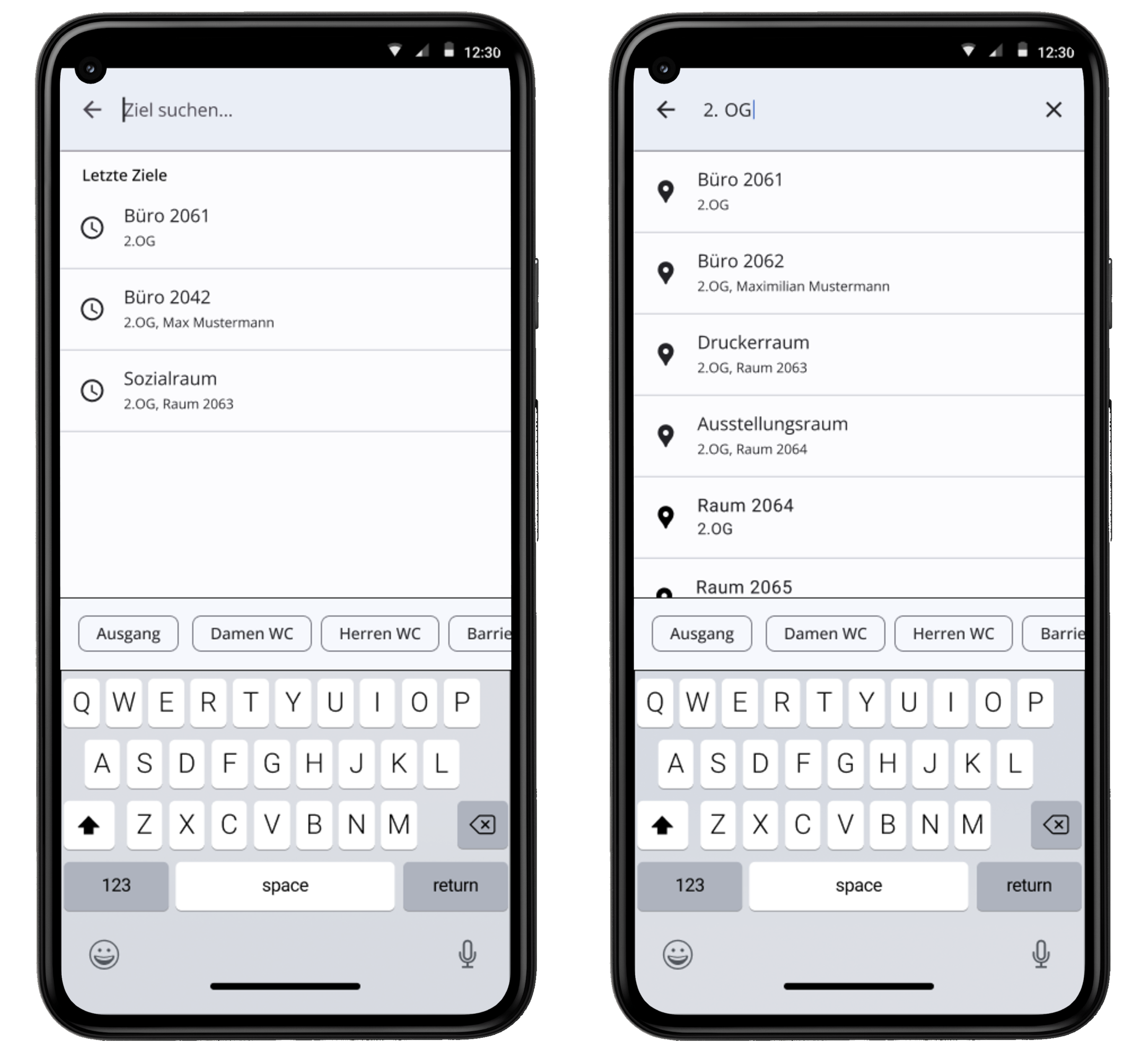}
    \caption{Search field (optimized for space and one-handed usage). Without search query, recent or frequent destinations are suggested (left), while adding a search query will filter the list of possible destinations (right).}
    \label{fig:search}
\end{figure}

People with visual impairments perceive colors and contrasts differently, e.g., due to color blindness. Designing an accessible app, the needs of all users must be considered. In the low-fidelity design phase, designs are first created in black and white to ensure that elements' differentiation and text readability depend on design features, rather than on color. Thereby, the impact of color perception issues is minimized.

The Material Design includes a theme with default values for the app's colors, typography, and other design elements. By adjusting the theme's colors, a custom color palette can be created and applied to all Material Design UI components. Moreover, the theme supports both light and dark modes. The Material Theme Builder tool is used to define this color palette. 

The colors in the palette meet basic contrast requirements but not at a standardized level for all users. Thus, the contrast of each background color and its associated text color is checked to achieve a minimum contrast ratio of 1:9. This ratio exceeds even the highest AAA level required by the WCAG standard, with colors that are less vibrant but more perceivable.

\subsection{Font selection}

A sans-serif font is used for good readability in the app. According to a project by the German Association for the Blind and Visually Impaired (DBSV) \cite{leserlich}, key points for readable text include distinguishable characters, character openness to avoid confusion (e.g., between 'c' and 'o'), consistent stroke thickness, normal character widths and spacing, and minimal bold or underlining. Open Sans, especially with the S3 variant for the letter 'I', fulfills these requirements excellently. The typography is defined in the Material Design Theme, largely keeping the theme's default settings but using Open Sans.

\subsection{Indoor navigation app structure}
As in the existing app, at first use, an intro is displayed, informing users about the functionality, handling, and adjustable navigation feedback options. Compared to the original version, this updated intro includes more detailed explanations.

After the intro, users access the main page displaying a map identical to the one in the existing app, allowing visual exploration of the environment without using the navigation function. Similar to other navigation apps, a search field on the main page leads users to the destination input page. To enhance one-handed usability, the search field is positioned at the bottom of the screen, considering the limited thumb reach for one-handed operation during navigation.

When users select the search field, they are directed to the destination input page featuring a focused search field and the device's standard keyboard, maintaining the familiar layout from the existing app. Below this field, users can access recent destinations quickly or filter through available destinations as shown in Fig. \ref{fig:search}. Predefined destinations are displayed as a horizontally scrollable list above the keyboard, optimizing space and enhancing one-handed usage.

\begin{figure}[hbt!]
    \centering
    \includegraphics[width=1\columnwidth]{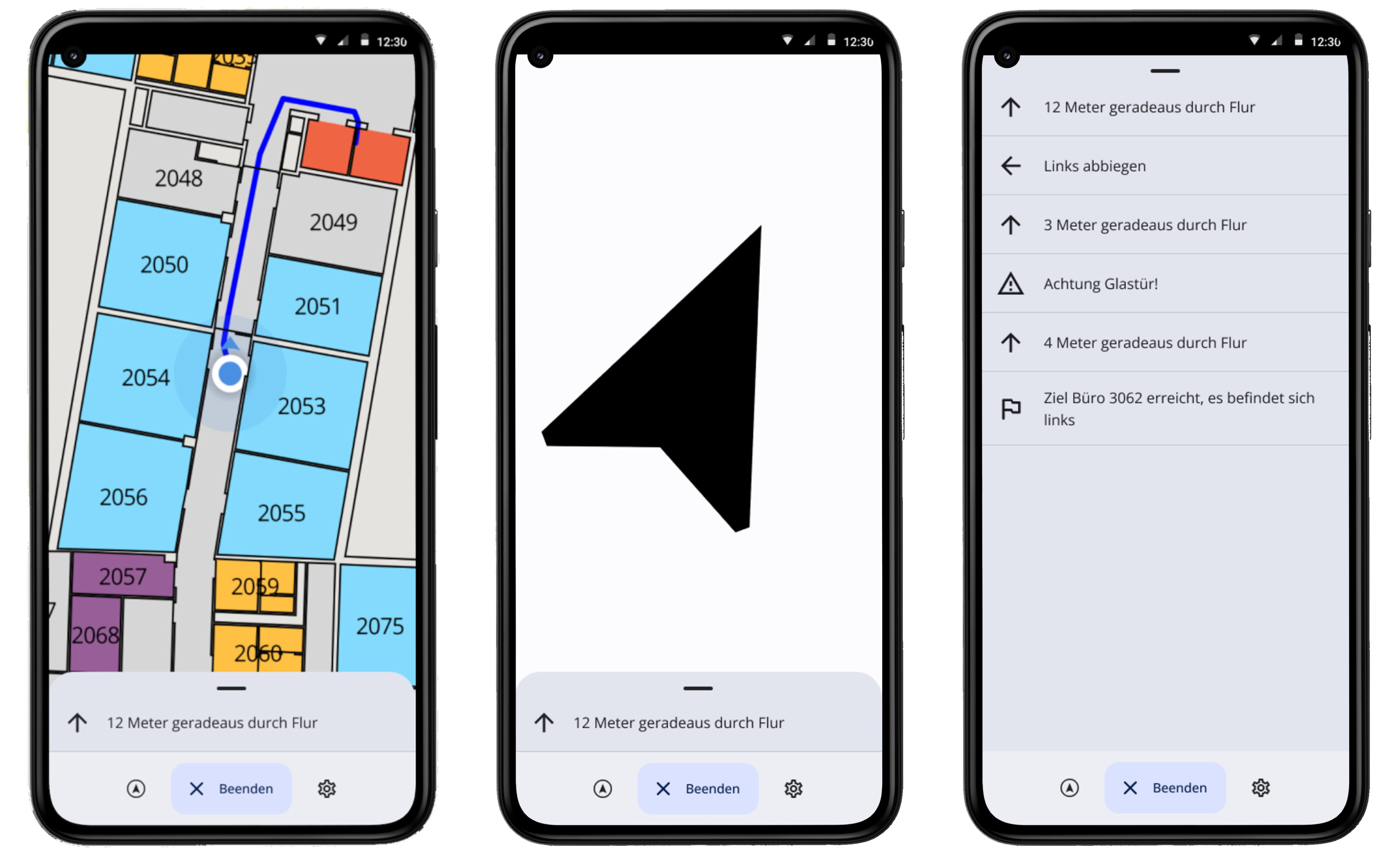}
    \caption{Navigation screens (left to right): Route on map, High-Contrast-Arrow pointing in the direction to go, Turn-by-Turn Instructions (opened by swiping the bottom sheet). The preview screens look identical, except not showing the current instruction.}
    \label{fig:navigation}
\end{figure}

Upon selecting a destination, users are directed to a preview page showing the route calculated from their current location. This feature allows users to familiarize themselves with the route without immediate navigation feedback. Two route display options are available: a visual map with the route marked and a textual step-by-step navigation list, both similar to Fig. \ref{fig:navigation}, the latter particularly useful for screen reader users.

For choosing the display method, various UI components can be used. Tabs are not suitable for one-handed operation as they are typically located at the top of the screen. An icon button, being more flexible in placement and space-saving compared to a text button, does not compromise readability with large text. However, for the primary action of starting navigation, a text button is preferred for its clear purpose recognition without needing to interpret icons. Both buttons are placed in a bottom bar for simple one-handed access to key local actions, with centered positioning for easy reach by both left and right-handed users.

\begin{figure*}[hbt!]
    \centering
    \includegraphics[width=0.75\textwidth]{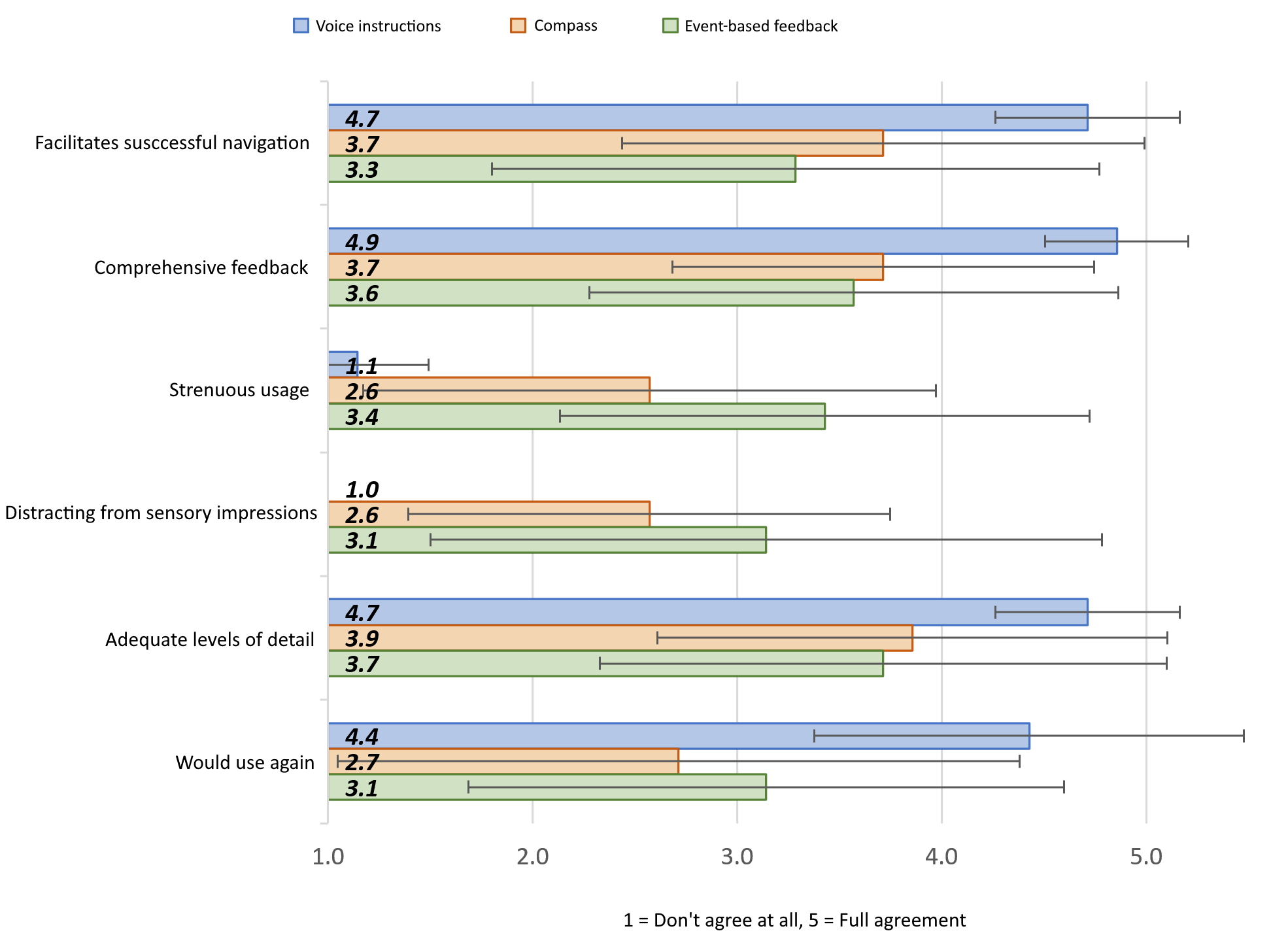}
    \caption{User study results: Voice instructions are unanimously rated positive with little variation, while compass and event-based feedback are subject to more variation because of personal preference. The acoustic and vibrating compass are evaluated jointly as well as the two event-based feedback concepts.}
    \label{fig:summary}
\end{figure*}

After starting navigation, the app displays a page with a map and a bottom bar for easy action access. The main control is a cancel button, with secondary options like an icon button to switch between map and arrow views, aiding visually impaired users with its minimalistic, high-contrast design. The navigation screens can be seen in Fig. \ref{fig:navigation}. An icon button gives access to the settings for adjusting navigation feedback. The navigation mode excludes the app header to adequately display map and arrow views, despite a bottom sheet showing the navigation steps above the bottom bar. This sheet, usually minimized to display only the current step, can be expanded to reveal the full route. The design contains a bottom bar spanning the screen, with clear icons and titles per entry.

\section{Evaluation}
\label{sec:evaluation}
In a user test, seven participants evaluate various methods of navigation feedback.
After completing the in-app tutorial for the respective feedback method, the participants are asked to use the app to navigate to a predetermined destination. This reference route had a length of 63m and included turns, an open space and several doors.

Initially, the general attitudes towards the three types of signals are surveyed. Voice advice is considered very useful for navigation feedback by nearly all participants (6 out of 7). In contrast, vibrations and audio signals receive mixed reviews. The specific voice prompts in the app are overall positively rated by all users. Many users value the short formulation of the advice without unnecessary fillers, enhancing clarity and ease of use. The results are shown in Fig. \ref{fig:summary}.

User evaluations of the compass concept vary which can partly be attributed to personal preferences. Generally, many users find the compass easy to understand with some practice. Most blind users report issues with the compass's high precision, especially on straight paths and near passage doors. This fine-tuning near doors distracts participants.

The survey asks participants about their preferred compass mode, either auditory or vibration. A clear majority of six participants prefers the vibration compass due to its perceived less intrusive nature with only one favoring the auditory version. Four participants opt for using the compass in combination with voice prompts for additional information. 

User evaluations of the event-based concept vary similarly to those of the compass concept, particularly regarding comprehensibility. Most users grasp the concept after some practice, as indicated by comments during practical tests, while others struggle to interpret the different impulses and patterns.

In practice tests, better comprehensibility of distance indications compared to directional instructions is observed. Some users suggest that increasing the gap between the success signal and directional pulses could significantly improve their perception.  Doors, in particular, confuse all users, as distance and success signals are communicated despite a straight path.

In the survey, users are asked about their preference for either the simple or complex version. Four participants prefer the simple variant, finding it somewhat more comprehensible.  Four testers also indicate that they would like to use the mode in combination with voice prompts. Sighted users appreciate the idea of an event-based feedback as an alternative to voice to reduce the need to look at the phone. 
In general, however, a reliable link between the type of impairment and the preferred feedback method cannot be observed due to the rather small test group and a suspected high impact of personal preference.

During practical tests, one issue is identified, affecting both the event-based vibration and the voice prompts. Participants navigate at very different speeds. As a result, some vibration impulses are skipped or played too late. Adjusting the walking speed resolved this issue almost completely. 

In the study, users with sufficient residual vision are also asked about UI design elements. Generally, they are rated very positive, contrast and font selection are considered appealing, and no problems with readability occurred during the user tests.

\section{Summary}
\label{sec:summary}

In this work we show how an accessible design of a smartphone app can significantly enhance usability for visually impaired users. The presented indoor navigation app aims to improve the independence and life quality of many people. A prototype implementation shows the potential of the proposed UI design. In this app, various direction concepts are realized. An acoustic compass guides users on the track with a click sound that gets faster the more the user drifts off. 

Vibration signals are an alternative concept for guidance. The implemented solution contains two different concepts for
direction encoding with vibration. The first option uses clock-like signals, e.g., two pulses for a direction of 2 o’clock, i.e.,
slightly right. The left and right direction are distinguished through the length of the vibration. The second method indicates a wrong direction with a single pulse. When the user turns, a signal for success is used, when the correct direction is reached. Distance signals are coded similar to distance warning systems in cars, i.e., short intervals for short distances and long intervals for long distances.

The user evaluation shows the guidance concepts' suitability. Especially the ease of use makes the event-based feedback useful for indoor navigation. Together with voice instructions that provide more detailed information, users are successfully guided to their destinations. The evaluation gives a good impression of the acceptance of the app, but tests with a larger user group are desirable. 
The users participating in the study used the app for the first time. A follow-up evaluation after a longer usage period could give more insight.

Additional work could consider the personalization of direction patterns. This can be achieved with settings or with pattern recognition, e.g., to detect the step length of individual users. Thereby, directions could be indicated with the number of steps instead of meters. Also, vibration patterns could be personalized, according to the preferences of the users.

\bibliography{paper}{}
\bibliographystyle{IEEEtran}

\end{document}